\documentclass[preprint,12pt]{elsarticle}

\usepackage{amsmath,amssymb,amsfonts}
\usepackage{algorithmic}
\usepackage{graphicx}
\usepackage{textcomp}

\usepackage{multirow}

\journal{Journal of Systems and Software}

\begin{document}

\begin{frontmatter}



\title{HPM-Frame: A Decision Framework for Executing Software on Heterogeneous Platforms}


\author[CTH]{Hugo Andrade\corref{mycorrespondingauthor}}
\cortext[mycorrespondingauthor]{Corresponding author}
\ead{sica@chalmers.se}
\author[CTH]{Ola Benderius}
\ead{ola.benderius@chalmers.se}
\author[GU]{Christian Berger}
\ead{christian.berger@gu.se}
\author[CTH]{Ivica Crnkovic}
\ead{vica.crnkovic@chalmers.se}
\author[CTH]{Jan Bosch}
\ead{jan.bosch@chalmers.se}

\address[CTH]{Chalmers University of Technology, Sweden}
\address[GU]{University of Gothenburg, Sweden}

\begin{abstract}
Heterogeneous computing is one of the most important computational solutions to meet rapidly increasing demands on system performance. It typically allows the main flow of applications to be executed on a CPU while the most computationally intensive tasks are assigned to one or more accelerators, such as GPUs and FPGAs. The refactoring of systems for execution on such platforms is highly desired but also difficult to perform, mainly due the inherent increase in software complexity. After exploration, we have identified a current need for a systematic approach that supports engineers in the refactoring process---from CPU-centric applications to software that is executed on heterogeneous platforms. In this paper, we introduce a decision framework that assists engineers in the task of refactoring software to incorporate heterogeneous platforms. It covers the software engineering lifecycle through five steps, consisting of questions to be answered in order to successfully address aspects that are relevant for the refactoring procedure. We evaluate the feasibility of the framework in two ways. First, we capture the practitioner's impressions, concerns and suggestions through a questionnaire. Then, we conduct a case study showing the step-by-step application of the framework using a computer vision application in the automotive domain.
\end{abstract}




\end{frontmatter}



\section{Introduction}
\label{sec:introduction}

The demands for performance of computer systems continue to increase, as the amount of functionality provided by software across multiple domains rises at a fast rate. The focus in most consumer electronics has been on software and its capabilities, resulting in a wide range of possibilities for users. Several complex features are achieved through techniques such as artificial intelligence (AI) using convolutional neural networks (CNNs), which require robust computation solutions due to high volumes of data to be processed \citep{Chen2017}. In the automotive domain, for instance, there are typically very high loads of sensor and camera data to be processed in real-time. Though computationally demanding, these systems are executed by embedded hardware with limited resources. It is often not feasible to significantly increase the computation power on-board due to constraints in terms of costs, weight, energy usage, and physical space.

One way to fulfil the aforementioned requirements is through heterogeneous computing, i.e., computer systems containing more than one type of processor, such as a combination of CPUs, GPUs and FPGAs. This solution allows different types of data to be processed in different ways. Typically, the main flow of the application is executed by the CPU, while the most computationally intensive tasks are executed by one or more accelerators. However, the adoption of this technology requires a number of changes to the typical software engineering processes, such as the inclusion of new tools, artifacts, and refactoring of existing code.
The latter is particularly relevant in the case of existing software that was initially designed for execution in a sequential format by CPUs. In some contexts, the portions of code to be accellerated are developed directly under a specialized framework for heterogeneous platforms \citep{Garland2008}. Further, despite the development of new technologies and frameworks within the community to facilitate software development for multiple hardware platforms, such as CUDA \cite{CUDA} and Vulkan \cite{vulkan}, the efforts in executing the required steps to successfully adopt such solutions cannot be neglected. 

\subsection{Research goal}
This paper aims at introducing a holistic approach that assists engineers in the task of refactoring software when heterogeneous platforms are introduced. Thus, the following research question is formulated: ``\textit{Which aspects are relevant when refactoring software for execution on heterogeneous platforms?}''

\subsection{Contribution}
The contribution of this work is a software engineering framework that aims at guiding engineers in the task of refactoring software in order to take advantage of the capabilities of heterogeneous computing. The decision framework consists of five stages to be followed. Each stage includes a set of questions to be answered in order to support the decision-making process for refactoring software in the aforementioned context. We have qualitatively evaluated the proposed framework through a questionnaire that was sent out to our industrial partners. The feedback also includes three interviews with software architects who are responsible for both research and production projects in industry. Additionally, we demonstrate the feasibility of the framework through a case study with an application in the automotive domain.

\subsection{Terminology}
In the context of this work, the term ``accellerator'' refers to a processor that is not the host processor in the hardware setup, typically the case of GPUs or FPGAs. The term ``processing unit'' refers to a computer processor, which is built based on one of the various types of microarchitectures. The term ``platform'' refers to a hardware component that contains one of more processors.

\subsection{Structure of the paper}
The remainder of this paper is organized as follows. 
We present the motivation in Section~\ref{sec:motivation}.
In Section~\ref{sec:researchmethodology}, we present the research methodology that was used. Section~\ref{sec:Case_Study} gives an overview of a system used as a case study in this work.
In Section~\ref{sec:framework}, we describe the proposed framework with a demonstration of its use on the aforementioned case study. We discuss a possible integration of the framework with a typical software development process in Section~\ref{sec:integration}. In Section~\ref{sec:evaluation}, we present the feedback from the practitioners. Then, we discuss the threats to validity in Section~\ref{sec:threatstovalidity} and the related work in Section~\ref{sec:relatedwork}. Finally, in Section~\ref{sec:conclusion} we conclude and present our plans for future work.

\section{Motivation---Example in the automotive domain}
\label{sec:motivation}
\begin{figure}[ht]
  \begin{center}
  \includegraphics[width=.75\linewidth,trim=0cm 0cm 0cm 0cm]{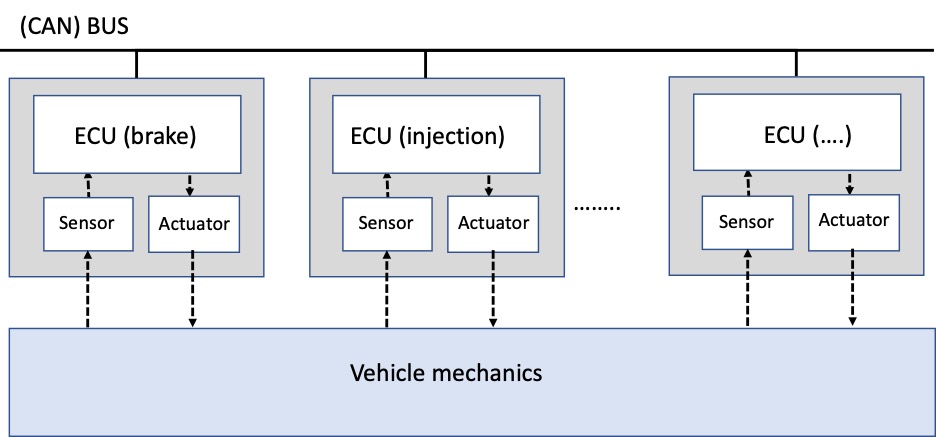}
  \caption{Example of a traditional vehicular control system architecture.}
  \label{fig:car-architecture}
  \end{center}
\end{figure}

Heterogeneous computing has the potential to bring benefits to systems in a variety of domains. In particular, the automotive industry provides an illustrative example of architectural evolution of the system deploying heterogeneous computing. In the last 25 years, the system and software architecture for vehicular control systems followed the best practices as emerged and shown in Fig~\ref{fig:car-architecture}. This architecture has satisfied the demands of the industry through simplicity over time, however it is currently undergoing changes in order to comply with the rapid increase in the number of functionalities that are embedded in modern products. 

The purpose of a system following the depicted model is to control different functions of a vehicle. The control system architecture is a distributed system consisting of many computational units (a.k.a. Electronic Control Units - ECUs) containing embedded software, typically including a control loop that receives signals from sensors, performing computation, and producing signals to the connected actuators that control the electromechanical parts of the vehicle. Typically, the communication between ECUs is realized through a standard CAN bus. Such modularized, component-based approach for the software features running on such ECUs is specified by the AUTOSAR standard \cite{AUTOSAR}. The standard focuses on evolution and scalability, allowing for the addition of new services through the introduction of new ECUs with embedded software. ECUs use simple CPUs and are dimensioned to maximally utilize the computational and memory resources that are automotive grade. 

Each ECU is designed using a layered architecture: (i) a hardware control layer connected to the sensors and actuators; (ii) a small real-time operating system; (iii) middleware for communication and standardized features for the high-level application logic; and (iv) a layer with concurrently-executing components that provide application functionality. 

An extension of this approach is building new functions and services in the form of distributed and shared software components. A component is distributed over several ECUs, and an ECU can contain several independent components. This implies that a component shares resources with other components in a particular ECU, and communicate synchronously or asynchronously over the bus. This approach decreases the number of ECUs, but requires a more sophisticated sharing of the resources on an ECU (CPU-time and memory, sensors and actuators), more communication resources, and more difficult resource optimizations which require more powerful CPUs and more memory (i.e. more expensive production). 

Several challenges are associated with this type of architecture, such as the strict constraints in fulfilling real-time requirements and the difficulties in implementing complex functionalities that require distributed components. In addition to controlling, the system has logging functions that are used for the analysis of the system performance. For this purpose, data is accessed either in real-time and sent to an attached device, or via small repositories that can be downloaded from the system in its non-operational state. A connection with the cloud usually occurs in a workshop environment where the vehicle is in \textit{testing mode}. 

In recent years, the development of new software and hardware technologies has enabled significant improvements in the automotive industry: in business models, in system performance, in safety, in impact on the environment, and in long-term development and maintenance costs. The main and disruptive changes are the transformation to electrical vehicles, autonomous driving, and connectivity. Examples of new functionality are: (i) different elements of autonomous driving; (ii) optimized engine control, (iii) improved behavior in risk-full situations; (iv) continuous monitoring or vehicle behavior; (v) continuous monitoring of driver's behaviour; (vi) substantial increase of infotainment functions; (vii) continuous cloud connectivity; and (viii) continuous deployment of new software versions.

Such disruptive change requires disruptive modifications in the system and software architecture, in the development and evolution processes, and to a large extent organizational changes. Being exposed to tough competition and low profit margins, the existing automotive companies apply a combination of continuous evolution of the existing products on one side, and start from completely new models, new architectures, and new organizations on the other side. 

Particularly in this domain, the new functions being introduced are typically computation-intensive, with extensive parallel computing, processing large amounts of data in real-time, and have major requirements on system performance. New technologies have appeared, including the use of machine learning (ML), parallel computing, intensive communication in real time, cloud computing and edge computing. Thus, a number of strategically important architectural decisions must be made in order to properly support business decisions and software development processes. 

\begin{figure}[ht]
  \begin{center}
  \includegraphics[width=.75\linewidth,trim=0cm 0cm 0cm 0cm]{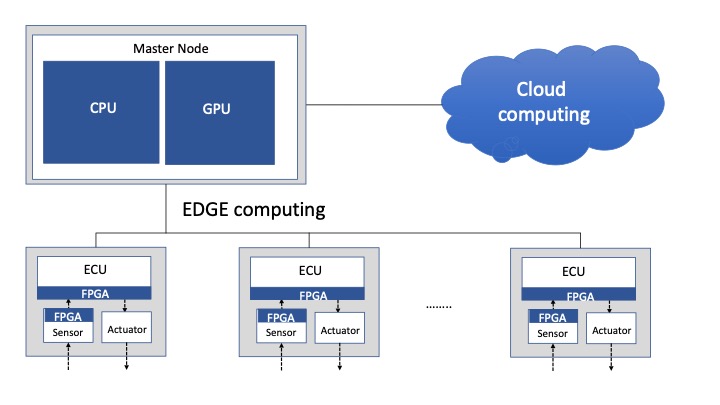}
  \caption{Example of a new system architecture for an automotive system.}
  \label{fig:new-car-architecture}
  \end{center}
\end{figure}

Fig~\ref{fig:new-car-architecture} shows a new architecture of an automotive system that provides architectural prerequisites for new functions and enables a continuous transformation from the old architecture to the new architecture. 

The basic architecture is the same---distributed systems connect via a bus, but the node structure is changing. Instead of nodes optimized for low computational and storage capacity, the nodes (ECUs) become heterogeneous computational platforms: (i) CPUs are getting more powerful, and in some cases replaced by multi-core CPUs; (ii) FPGAs are being included on the platforms for specialized computation, in particular processing input data from sensors (such as camera or radars). Further, the sensors are equipped with computational platforms (typically CPU + FPGA), enabling direct data processing and significant reduction in the amount of data that is sent for further processing. Additionally, the computing power is concentrated on a new, centralized, powerful computational platform that includes (multi-core) CPUs and GPUs, and in this way, many functions from distributed ECUs can be moved to it. Thus, the most-intensive computational services can be performed in real-time. This computation altogether can be seen as \textit{edge computing} in respect to the cloud computing to which the automotive system is connected, but not necessarily continuously. The cloud computing resources are used for additional services and services that do not have hard real-time requirements. Further, the cloud computing resources are used for further development of the system, including the training of machine learning models, and analysis of the data provided by the monitoring and logging functions of the vehicles.

The aforementioned example shows the complexity of the process, which raises important architecture-related questions. Some of them, related to software deployment on heterogeneous platforms, are listed next. 
\begin{itemize}
\item Deployment process-related decisions:
\begin{itemize}
    \item What is the process of migration of the system architecture? 
    \item What is the process of migration of existing code from a platform (CPU) to a heterogeneous platform (executable on a e.g., GPU, or FPGA)? 
    \item How to apply continuous deployment of software on the changing system architecture?
    \item What are the implications of deployment of new architectures on the overall system's properties (resource utilization, like energy consumption, overall system performance, development and production costs, etc.)?
\end{itemize}
\item Deployment of a new architecture:
\begin{itemize}
    \item How to transform the existing architecture into the new architecture? 
    \item How to distribute services in the new architecture, and ensure quality properties related to real-time requirements, performance, resource utilization, etc? 
\end{itemize}
\item Software deployment:
\begin{itemize}
    \item How to transform the existing architecture into the new architecture?
    \item How to distribute services over the heterogeneous computational units and ensure quality properties related to real-time requirements, performance, resource utilization, etc? 
\end{itemize}
\end{itemize}

These questions have many sub-questions and require many decisions of different types. Additionally, the decisions are directly and indirectly inter-dependent. For this reason, it is very challenging to successfully manage the deployment process, and there is a high risk that non-coordinating decisions lead to major problems, such as infeasible or poor solutions, and increased development and maintenance costs. 

Based on the aforementioned topics, and considering the level of complexity involved in the process of migration, a systematic approach that defines the complete decision process and its implementation is needed. 

\section{Research Methodology}
\label{sec:researchmethodology}

\begin{figure}[ht]
  \begin{center}
  \includegraphics[width=.85\linewidth,trim=0cm 0cm 0cm 0cm]{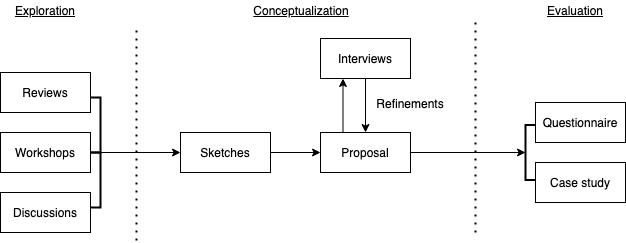}
  \caption{Overview of the research stages that were conducted.}
  \label{fig:methodology}
  \end{center}
\end{figure}

This work is part of a series of studies in the area of heterogeneous computing. It was conducted with basis on the design science methodology \citep{Dresch2014} (or \textit{constructive science research}), which aims to establish and operationalize research when the desired goal is an artifact or a recommendation. The application of design science research culminates in new ideas or a set of analytical techniques that enable the development of research \citep{Vaishnavi2007}. As shown in Figure~\ref{fig:methodology}, the proposal presented in this paper was created based on two different perspectives: the state-of-the-art and the experiences of practitioners in large industrial contexts.

The exploration stage consisted of studying the available literature on the topics of software deployment \citep{Andrade2019} and software architectures \citep{Andrade2018} for heterogeneous platforms in order to obtain an overview of the research area and identify the challenges, concerns, and gaps in research. In these studies, we identified several approaches to realize software architecture design and deployment when a heterogeneous platform is available. 

In the conceptualization stage, we began by sketching a solution based on the knowledge gathered during exploration. We iteratively identified the gaps between theory and practice through discussions with multiple partners in industry, capturing their practices, and analyzing their perspectives in trying to meet different challenges \citep{Andrade2019b}. The sketches were discussed and refined within the group of researchers. When it had reached a specified level of maturity, we scheduled face-to-face meetings with our industrial partners. These meetings were conducted in the form of interviews or workshops, in which we presented the framework and discussed the topics. Both positive and negative aspects were raised and captured through annotations. 

Finally, we evaluated the approach through a questionnaire that was sent to several companies, including the ones that participated in the workshops. This procedure was designed to capture the practitioners' impressions of the framework through explicit, open questions about their daily experiences on the topic. The questions are strongly tied with the concerns and questions proposed in the framework, in order to obtain feedback that is specific to each stage. The responses were qualitatively analyzed and the highlights of the feedback are presented later in this paper.

\section{Case Study}
\label{sec:Case_Study}

The case study was built in the context of the open-source software platform OpenDLV for self-driving vehicles. We have used it to demonstrate the activities defined in the proposed framework for its deployment on a heterogeneous platform. In this section, we provide a brief overview of the case study, as follows.

OpenDLV \cite{opendlv} is an open-source software platform that provides the necessary software infrastructure for applications in the automotive domain with a strong focus using the microservice-based architectural paradigm and continuous deployability. The platform is actively used at the Chalmers laboratory Revere \cite{Revere}, a research facility part of Chalmers University of Technology that focuses on autonomous systems, active safety, vehicle dynamics, and large-scale data processing for ML-based vehicular applications.

OpenDLV supports the development and testing of applications in the autonomous driving domain through an architecture based on microservices. It provides transparent means for communication between software components running on distributed nodes including automotive applications for sensor fusion and autonomous driving. One of the goals of OpenDLV is to abstract the complexity of lower layers of computation and separate them from the application logic, offering developers a standardized message passing protocol that is implemented transparently exploiting the computational and network features at hand.

Since it is entirely based on microservices, there is a strong focus on deployment, where in the typical case each microservice is packaged and shipped using Docker images \cite{Docker}. A Docker container runs inside a Linux namespace and includes all necessary dependencies for the execution of functionalities, therefore allowing for an isolated runtime environment. Furthermore, to automate the development and deployment process, the microservices are built automatically in build pipelines made available via GitLab.

OpenDLV is realized with libcluon, a single-file, header-only library for portable C++ microservices \cite{libcluon}. All message-passing functionalities such as network-based communication (TCP and UDP) and shared memory for same-node communication are provided by the library can be obtained through importing the library into a C++ project and compiling it with a modern C++ compiler.  

In this case study, a simple edge detector implemented as a 2D convolution operation will be investigated using the framework. The per-frame operation can be heavily parallelized since per-pixel operations are completely independent for convolution. A pre-recorded video stream of 640x480 pixels is used in the demonstration, where only the red color channel is used. For the target system, a single channel image stream is envisioned. The convolution kernel is an edge detector 3x3 matrix defined as:
\vspace{0.3cm}
\[
K=
  \begin{bmatrix}
    0 & 1 & 0 \\
    1 & -4 & 1 \\
    0 & 1 & 0
  \end{bmatrix}
\]
\vspace{0.3cm}

\begin{figure}[ht]
  \begin{center}
  \includegraphics[width=1\linewidth,trim=0cm 0cm 0cm 0cm]{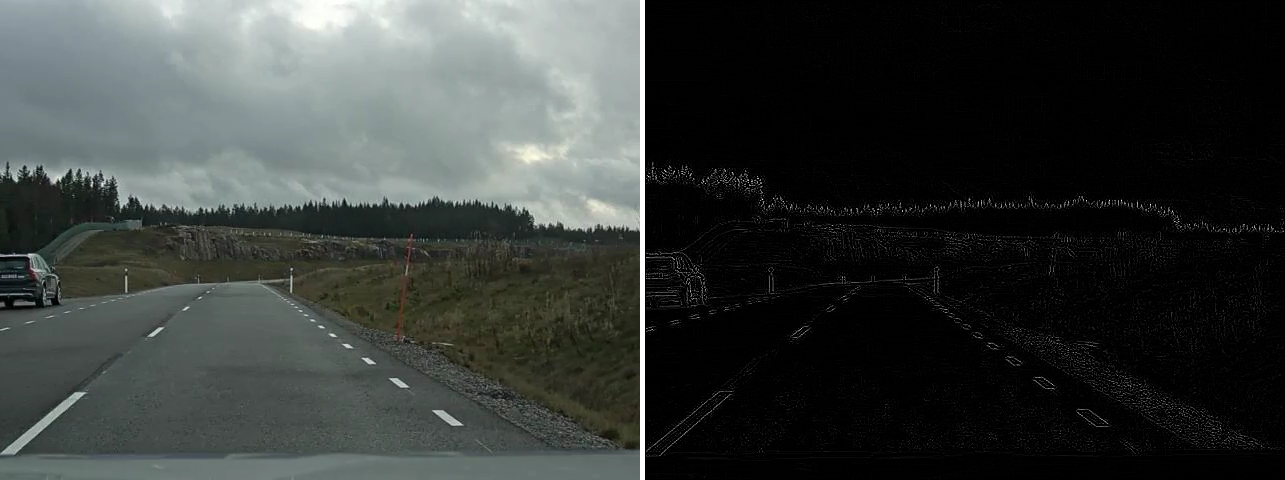}
  \caption{An example input and output from the 2D convolution considered in the case study.}
  \label{fig:usecase}
  \end{center}
\end{figure}

A CPU version of the 2D convolution was implemented as a simple OpenDLV microservice was used as a starting point, and the proposed framework was used for the purpose of bringing this implementation into a GPU-based microservice in a heterogeneous system \cite{CaseStudyRepo}. The result of the 2D convolution algorithm is shown in Figure~\ref{fig:usecase}.

\section{The HPM Framework}
\label{sec:framework}

The HPM (\textbf{H}eterogeneous \textbf{P}latform \textbf{M}igration) process framework provides software practitioners with aid in the process of re-architecting software systems (or including new features) when introducing heterogeneous hardware platforms. Focusing on the software engineering perspective, HPM supports practitioners in the process of effectively migrating CPU-centric applications into software that can be executed on heterogeneous platforms. The HPM framework is an extension of our previous work focusing on the re-architecting stage \citep{Andrade2020}.

\begin{figure}[ht]
  \begin{center}
  \includegraphics[width=.85\linewidth,trim=0cm 0cm 0cm 0cm]{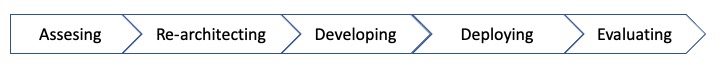}
  \caption{Stages of the HPM framework.}
  \label{fig:hpm}
  \end{center}
\end{figure}

As shown in Fig~\ref{fig:hpm}, the framework is defined in the form of a process consisting of five stages with the following characteristics:

\begin{itemize}
\item [-] \textit{Assessing}: reasons about the feasibility of system refactoring towards heterogeneous platforms;
\item [-] \textit{Re-architecting}: provides the new architecture of the platform or software, deciding which components will run on which execution platform;
\item [-] \textit{Developing}: provides detailed design and implementation of code;
\item [-] \textit{Deploying}: determines delivery policies and prepares the target hardware;
\item [-] \textit{Evaluating}: attests functionalities and evaluates the refactoring process.
\end{itemize}

While these stages are differentiated by their goals and methods, they are not necessarily performed separately or sequentially. They can, for example, be performed in parallel and in several iterations, depending on the adopted software development process within the entire application lifecycle. Each stage includes a number of main concerns that should be addressed and questions that should be answered in order to obtain decision support for the refactoring procedure.

In the next subsections, we present the stages in more detail, along with the topics that should be acknowledged and the questions that should be answered. We then demonstrate application of the framework by discussing each topic and answering the questions in the context of our OpenDLV application.

\subsection{Assessing}
The goal of this phase is to \textit{assess the feasibility, advantages, disadvantages, and preliminarily the impact of refactoring}. 

At this stage, the bottlenecks of the system must be identified with the intention to improve a certain non-functional property, such as performance. Then, one should assess the possibility of redesigning the identified portion by highlighting the constraints in doing so. Afterwards, calculations and simulations should be conducted in order to determine the expected benefit of re-designing. Once the technical feasibility is assessed, one can determine the type of hardware to be used. Discussions about the required expertise and impact of the changes on the system must also be conducted. 

The main concerns of this stage are:

\begin{enumerate}
\item Analyzing the functionality to be executed on the accelerator(s);
\item Determining the hardware to be used;
\item Determining the impact on the overall system and stakeholders;
\item Determining the feasibility of the project.
\end{enumerate}

\subsubsection{Analyzing the functionality to be executed on the accelerator(s)}
The need for improved performance must be identified through the identification of functionalities to be accelerated. The software that implements the identified functionalities must be assessed in order to determine whether or not it is parallelizable. The need for improved non-functional requirements must be identified---being, for instance: performance or energy efficiency. There must be an evaluation of the existing system in the search for bottleneck(s) to be analyzed in terms of (i) extension, (ii) risk level (to the business); and (iii) urgency level. The portions of the system that require improved performance should be highlighted and analyzed.

We address these aspects through the following questions.

\vspace{0.5cm}
\noindent\fbox{%
    \parbox{\linewidth}{%
        Q1: Are there bottlenecks in the system?

        Q2: Which parts of the system require better performance? 

        Q3: Would the selected function perform more efficiently on a different type of processing unit?
        
        Q4: Is the existing software parallelizable?
    }%
}

\vspace{0.5 cm}
\textit{The case study:} The 2D convolution cannot be fully parallelized using a CPU (Q1). Each per-pixel convolution operation is completely independent and full parallelization could therefore be reached with an appropriate accelerator (Q4). The 2D convolution is the main per-frame operation for the edge detection algorithm (Q2). Therefore, increased computational performance for this operation has a direct connection to the data output rate of the software component. An accelerator that could operate on each pixel of the image frame in parallel would greatly increase the performance of the software component (Q3). The existing CPU-based software run per-pixel operations in sequence, for each frame. However, there are 307,200 pixels in each input image, so a CPU-based embedded system can not efficiently run the operations in parallel (Q1 and Q2).

\subsubsection{Determining the hardware to be used}
Once the feasibility of refactoring is assessed with regards to the software, one must determine the hardware that will be used. There exist several options that are initially designed to perform a computation on a specific type of data. There are also other options that focus on general-purpose architectures. A hybrid solution between the two is also possible through, for instance, using a GPU as a general-purpose computing platform. The decision about the hardware also affects software development processes and tools that must be used, such as manufacturers' provided frameworks and APIs.

We address these aspects through the following questions.

\vspace{0.5cm}
\noindent\fbox{%
    \parbox{\linewidth}{%
        Q5: What type of hardware should be used?
        
        Q6: What hardware vendor should be used? 
        
        Q7: What vendor-specific toolchain is needed for the change?
    }%
}

\vspace{0.5 cm}
\textit{The case study}: Since the operation on each frame can be fully per-pixel parallelized, and since the 2D convolution is a simple operation, an accelerator that is designed to run a large number of small operations in parallel is preferred. Therefore, a GPU was selected as an accelerator (Q5). Furthermore, to allow for good deployability and hardware-scalability, the Vulkan API was selected as the means of interaction between the co-processor (CPU) and the GPU (Q6). Since Vulkan is an open industry standard, no vendor-specific tools are needed, and as the software in this case is simple, pure SPIR-V can be used, either hand-coded or via GLSL (Q7).

\subsubsection{Determining the impact on the overall system and stakeholders}
The impact of the changes should be preliminary assessed in a way in order to determine the expected benefit with the intended refactoring of the system. Potential changes in the architecture should be carefully analyzed. Particularly, the communication between components should be revised in order to foresee the results of the changes. Ideally, the gains in performance should reflect on the perceived non-functional requirements by the users. 

We address these aspects through the following questions.

\vspace{0.5cm}
\noindent\fbox{%
    \parbox{\linewidth}{%
        Q8: What would be the impact on the existing architecture?

        Q9: What would be the impact on the system's performance and lifecycle?

        Q10: What are the expected business benefits (time-to-market, market advantage, user perception)?
    }%
}

\vspace{0.5 cm}
\textit{The case study}: 
Since the GPU needs a co-processor, no changes to the overall architecture are foreseen, as the interface from the camera sensor to the new implementation is unchanged as well as the resulting output to other original components (Q8). As a result of the refactoring, the performance is expected to be greatly increased, better matching a higher input frame rate, into the output frame rate. The scalability of the solution is also simple, as embedded GPUs from different vendors can be used without changing the Vulkan and SPIR-V implementation (Q9). Furthermore, the main business benefit with the change is that the organization can more easily adapt to future changes in hardware outside this particular component (Q10). If a camera of higher resolution, or higher frame rate, a GPU accelerated component can easily be integrated to match the data input stream, without any vendor-constraints or software changes.

\subsubsection{Determining the feasibility of the project}
If new functionalities are to be developed, there must be an analysis on their requirements in order to determine the feasibility of executing them on a heterogeneous platform. In some cases, there exist heavy constraints for the re-design of the system, such as increased complexity, legacy codebase, and lack of resources. Furthermore, project-related aspects should be assessed, such as the required expertise, and the cost of re-architecting the system. In particular, the human resources (expertise) should be planned, as well as the cost of implementing the changes, with the goal of obtaining a favorable return of investment.

We address these aspects through the following questions.

\vspace{0.5cm}
\noindent\fbox{%
    \parbox{\linewidth}{%
    Q11: What would be the cost of the changes?

    Q12: What would be the costs to maintain and evolve the functionalities once migrated? 

    Q13: What would be other consequences of the change regarding non-functional properties?
    }%
}

\vspace{0.5 cm}
\textit{The case study}:
Since Vulkan and SPIR-V are open standards, and no special tooling is needed, no further vendor contracts or tooling acquisitions are needed (Q11). As the code itself is simple, it can mostly be included in the current CI/CD environment. Possibly, to fully allow for tests, a hardware-in-the-loop (HIL) system might be needed in the backend, but also this is vendor-agnostic and can be freely configured by the organization. Since no special vendor relations are needed, no large increase in costs are foreseen. However, some internal training on SPIR-V and Vulkan might be needed, but due to the simple nature of this particular implementation no special demands on such training is expected (Q12). The solution will bring market advantage by allowing for improved performance of the computer vision algorithms (Q13).

\subsection{Re-architecting}
The goal of this phase is to \textit{define the necessary changes in the architecture}.

One of the intrinsic activities of implementing software for heterogeneous platforms is determining which portions of software will be executed by which processor. Such setup might heavily influence on the overall architectural design, allowing or demanding for particular patterns and principles to be used. The new architecture should then be aligned with other surrounding aspects in the project.

The main concerns of this stage are:

\begin{enumerate}
    \item Mapping software and hardware;
    \item Determining the overall architecture design;
    \item Determining the impact on the new architecture.
\end{enumerate}

\subsubsection{Mapping software and hardware}
This activity consists of determining which portions of software will be executed by which hardware processors. This decision is particularly sensitive due to the great difference in performance that can be observed depending on the chosen software/hardware configuration. In summary, the engineers should determine the preliminary performance of potential configurations (e.g., using tools embedded in common vendor frameworks), and use the results to decide on the most appropriate configuration.

We address these aspects through the following questions.

\vspace{0.5cm}
\noindent\fbox{%
    \parbox{\linewidth}{%
        Q14: Which functionalities will be executed by the accelerator(s)?
        
        Q15: How do the potential configurations perform? 

        Q16: What will be the mapping between software components and processing units?
    }%
}

\vspace{0.5 cm}
\textit{The case study}: 
The embedded system that needs to run the new implementation needs to be a combination of co-processor and GPU. The CPU needs to fetch, upload, and download image frames to the GPU, and the GPU needs to carry out the 2D convolution (Q14). The fetching of the video image can be unchanged from the original implementation. This is very hardware-dependent and can be tuned based on current and future needs. For the current solution 307,200 operations are needed for each frame, and a GPU suitable for embedded systems may run a few tens of thousand threads in parallel, meaning that less than 30 iterations per frame is expected (Q15). The current CPU-based system runs one operation in about 16 ns, with a total mean time of 4.8 ms per frame. Therefore, we could expect the accelerator to reduce the per-frame operation from 4.8 ms to 480 ns. However, as the frame also needs to be transferred to and from the accelerator for each iteration time for this needs to be added on top of the pure computation time.  For the new setup, two software parts are needed: the CPU software to interact with the Vulkan API to utilize the GPU device and memory, and the SPIR-V software that is at runtime turned into GPU machine code and then uploaded by the CPU via the GPU driver (Q16).

\subsubsection{Determining the impact on the new architecture}
When the overall architecture is determined, and possible mappings between software and hardware are defined, it is important to analyze the concrete impact on the new architecture, not only in performance but also in terms of other quality attributes. The impact of these changes should be analyzed in order to plan for a design that is most appropriate.

We address these aspects through the following questions.

\vspace{0.5cm}

\noindent\fbox{%
    \parbox{\linewidth}{%
        Q17: What is the impact of the new architecture on the constraints and requirements at run-time?

        Q18: What is the impact of the new architecture on the constraints and requirements related to lifecycle and business aspects?

    }%
}

\begin{figure}[!ht]
  \begin{center}
  \includegraphics[width=.6\linewidth,trim=0cm 0cm 0cm 0cm]{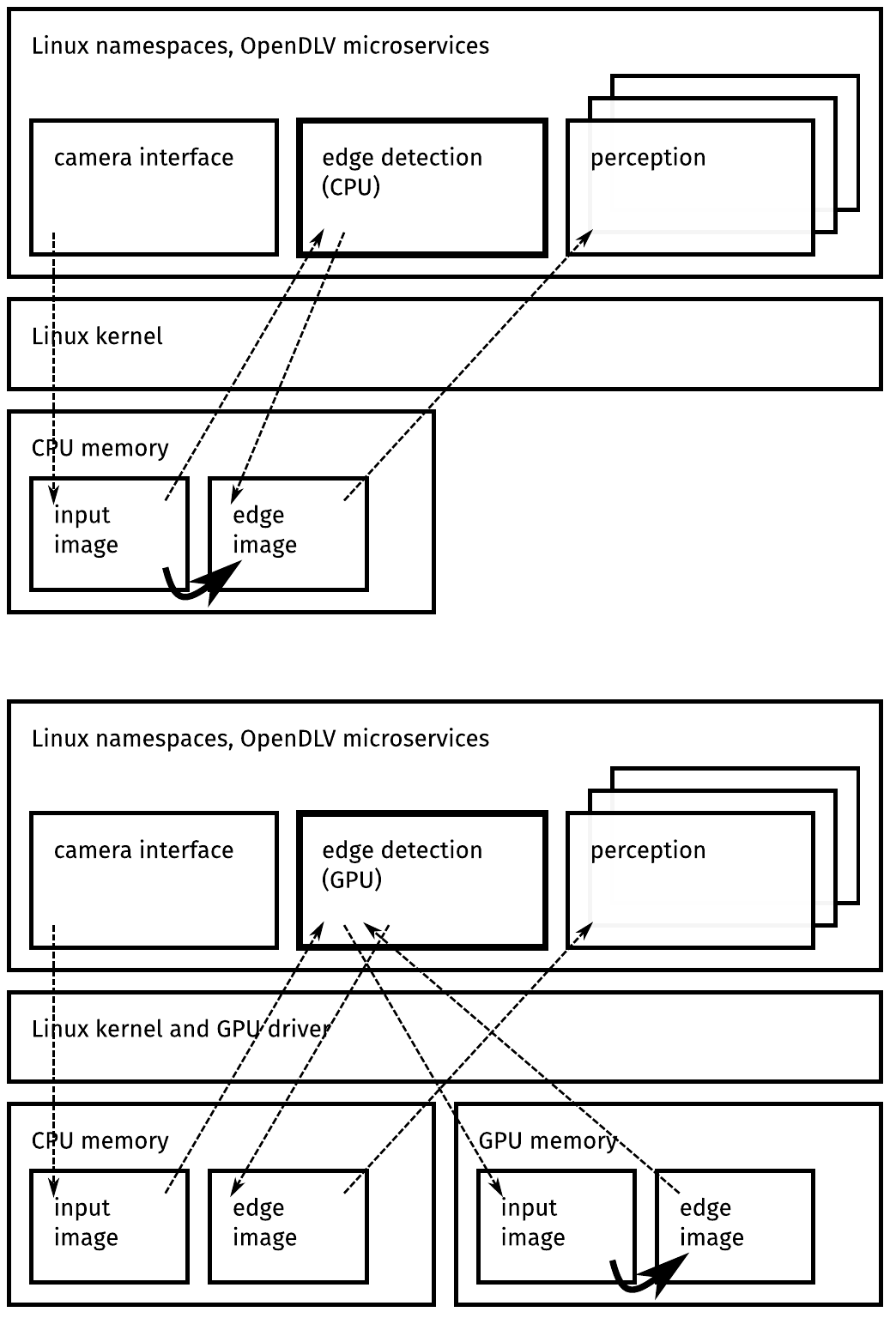}
  \caption{System overviews connected to the two implementations in the OpenDLV use case. The microservice under study is marked with strong borders. The top panel shows the initial CPU implementation, and the bottom shows the resulting CPU and GPU version. The dotted arrows show memory flow and should be read from left to right, and the strong arrow indicate the calculation step.}
  \label{fig:opendlv-usecase}
  \end{center}
\end{figure}

\vspace{0.5 cm}
\textit{The case study}: 
A number of architectural attributes are modified from the CPU-based to the GPU-based implementation, as shown in Figure~\ref{fig:opendlv-usecase}. The data flow of the original CPU implementation is defined as follows. The input image is obtained from the camera interface and stored in the CPU memory. Then, the edge detector function---implemented by an OpenDLV microservice---is executed on the CPU, provided an image based on the identified edges. Such result is then provided to the perception service. In the GPU implementation, the edge detector function is executed by the accelerator using the image stored in the GPU memory space and the calculation step results in the edge image. Both images are stored in the GPU memory space (Q17).

The performance requirements of this system are connected to three main aspects: faster execution time, modularity, and maintainability. The edge detector service is expected to perform faster when compared to the CPU implementation, with the goal of approaching the capacity of the camera in terms of frame rate as much as possible. The modularity and maintainability of the system are achieved by using the microservices-based approach to software development, regardless of the underlying hardware technology that is used. It allows microservices to be individually created and maintained. Faster edge detection results in faster response of the overall system in providing real-time computer vision functionality (Q18).

\subsubsection{Determining the overall architecture design}
The introduction of a heterogeneous platform and consequently determining a software/hardware configuration may require changes in the overall software architecture. There are several architectural patterns and architectural design principles that can be used in the context of heterogeneous platforms, mainly addressing the communication and computation aspects.

We address these aspects through the following question.

\vspace{0.5cm}
\noindent\fbox{%
    \parbox{\linewidth}{%
        Q19: What will be the new/changed architectural structure, components, and communication?
    }%
}

\vspace{0.5 cm}
\textit{The case study}: 
The overall architecture will not change, as the accelerator is managed with a co-processor that runs the original interfaces. All communication is handled by OpenDLV, including input and output of signals and messages. On the software side, the communication is handled by a standard message set (SMS) protocol, which decouples the high-level application logic from the low-level device drivers through standardized messages. Due to the microservices-based architecture, the system is extensively modular, allowing for reuse of the same component in new vehicle models, where only the GPU needs to be upgraded to meet the new requirement specification (e.g. new camera frame rate or resolution). In this case, some characteristics from other patterns can also be observed, such as the primary/secondary pattern, in which the main processor dispatches a particular task to the accelerator and waits until the execution is finished before resuming the application flow (Q19).

\subsection{Developing}
The goal of this phase is to \textit{implement the solution (code) and realize changes in the system architecture}. 

At this stage, one must establish a development process that best suits their current needs and using the manufacturer's supported development frameworks. For instance, in the case of NVIDIA GPU programming, the CUDA framework is typically utilized for software development. It is also important to acknowledge eventual platforms' limitations. In the case of software, there are typically programming language requirements. In terms of hardware, there are often limitations in terms of the type of data that will be executed by the processing unit.

The main concerns of this stage are:

\begin{enumerate}
\item Adopting a software development process;
\item Refactoring software components.
\end{enumerate}

\subsubsection{Adopting a software development process}
Developing software for different computing units requires new programming languages, testing techniques and technologies to be incorporated. Most importantly, the adoption of a new software development process that supports a range of requirements is needed.

We address these aspects through the following questions.

\vspace{0.5cm}
\noindent\fbox{%
    \parbox{\linewidth}{%
        Q20: What are the limitations of the existing tools, frameworks and programming languages?

        Q21: What tools, frameworks and programming languages will be used?

        Q22: What changes will affect the development process (development, testing, etc)?
    }%
}

\vspace{0.5 cm}
\textit{The case study}:
The software development process includes the Vulkan API, SPIR-V, and C++ (Q21). The new implementation would not be possible solely with C++ due to hardware architecture constraints (Q20). Since the input and output of the microservice are the same, and since the co-processor needs to handle these interfaces, parts of the CPU code could simply be carried over to the new implementation (Q20). Specifically, the code that reads the shared memory populated by the camera microservice could be kept unchanged, and the code that saved the resulting image from the 2D convolution could be kept the same. 

In terms of tools and frameworks, the decision was made based on the type of hardware that was selected (Q21). Currently, there are strict restrictions in using the CUDA development framework with non-NVIDIA hardware. The manufacturer's business strategy seems to be related to the inter-dependence of their provided software-hardware solutions. Although there is vast documentation and support forums online for software development for GPUs using CUDA. However, both the development tools and the required drivers are relatively heavyweight in complexity and file size. The latter represents a threat to the automotive domain, in which deployable binaries must be as lightweight as possible, running on very resource-constrained hardware. For this reason, and due to the possibility of using different hardware platforms, Vulkan was adopted as the development framework. The development and testing of the modules must comply with the syntax and procedures of the Vulkan framework (Q22).

\subsubsection{Refactoring software components}
Component refactoring includes rewriting parts of the code to be executable on another computing unit. Important topics to sort our include: selection of technology, decision about having components available on multiple computing units, or having code for one computing platform, how data is processed between the platforms and, how the interface between components are specified, and similar.

We address these aspects through the following questions.

\vspace{0.5cm}
\noindent\fbox{%
    \parbox{\linewidth}{%
        Q23: What type of interface will be defined for the heterogeneous components?

        Q24: How will data be transferred between the platforms?

    }%
}

\vspace{0.5cm}
\textit{The case study}:
The change was only related to the actual operation, where the original CPU version directly operated on a copy of the incoming memory into a copy of the output memory (Q23). The new implementation instead copies the shared memory into the GPU memory and initiates a compute shader to generate the result. As the last step, the results are copied from the GPU directly into the output shared memory for further use in other vision algorithms (Q23 and Q24).

\subsection{Deploying}
The goal of this phase is to \textit{determine characteristics that are relevant to deployment, including the preparation of the target hardware}. 

The deployment scheme should be planned in a way that the expected non-functional requirements are met while the execution capabilities of the processing units are taken advantage of. Further, the decisions made during stage 2 ``Re-architecting'' should be reflected here. In particular, communication aspects should be carefully implemented and tested due to the inherent properties of heterogeneous computing. The computation should be parallelized whenever possible, in order to optimize the execution of functions on the heterogeneous hardware. Furthermore, the target architecture should then be prepared for execution, e.g., the appropriate drivers should be installed prior to execution.

The main concerns of this stage are:

\begin{enumerate}
\item Determining a deployment process;
\item Preparing the target hardware.
\end{enumerate}

\subsubsection{Determining the deployment process}
Crucial aspects to deployment include the integration of new functions and the update of existing functions. Changes can either be introduced \textit{one-by-one} (i.e., incrementally) or in batches. The limitations of the resources play an important role in this stage, i.e., for a resource constrained embedded system based on FPGA, it can be prohibitive to deliver smaller functionalities one at a time. Further, issues related to communication and bandwidth may arise due to the limitations of the hardware. In any case, the functional integrity of the system must be preserved despite the changes in the codebase.

We address these aspects through the following questions.

\vspace{0.5cm}
\noindent\fbox{%
    \parbox{\linewidth}{%
Q25: Will deployment be available only in a development phase, or will dynamic deployment be available?

Q26: Which constraints exist in the deployment process (timing constraints, resource capacity, and similar)? 

Q27: Can components be deployed independently, or all components must be (re)deployed as a set?

Q28: What are the capabilities of saving data and system status on the target processing units?
    }%
}

\vspace{0.5cm}
\textit{The case study}:
The co-processor is generating and uploading GPU machine code at runtime using the SPIR-V binary, so re-deploying software only needs a system reboot. The GPU and co-processor performance is likely scaled towards a specific camera system (fixed frame rate and resolution), but minor adjustments to the software might be possible in new software deployments (Q25). The SPIR-V binary can potentially be deployed independently, but it was decided to make the SPIR-V as an in-line specification in the co-processor software in order to maintain a atomic contract between the two components (Q27). Since the system is reactive, limited work memory is needed, nor included. Both the co-processor and the GPU needs about 617 kB of memory (Q26 and Q28).

\subsubsection{Preparing the target hardware}
The target hardware and software platforms must be prepared prior to executing the re-architected application(s). The deliverables should also be subject of analysis since the different compilers generate files of different types for execution across the heterogeneous hardware.  

We address these aspects through the following question.

\vspace{0.5cm}
\noindent\fbox{%
    \parbox{\linewidth}{%
        Q29: What are the preparations needed for execution?
    }%
}

\vspace{0.5cm}
\textit{The case study}:
In this case, no special preparations are needed, as the GPU machine code is generated at runtime (Q29).

\subsection{Evaluating}
The goal of this phase is to \textit{evaluate the solution and analyze the outcomes of the process}. 

After deployment, there should be a testing mechanism to attest that the expected benefits were achieved with the re-design. Also, the old functionalities should be tested in order to ensure that they were preserved. Finally, there should be an evaluation of the cost-benefit involved in the process of refactoring. Typically, one would assess all cost involved in the process against the gains in non-functional properties, such as performance or energy consumption.

The main concerns of this stage are:

\begin{enumerate}
\item Ensuring that the functionality has been maintained;
\item Determining the outcomes of refactoring.
\end{enumerate}

\subsubsection{Ensuring that the functionality has been maintained}
After multiple changes to the codebase, there must be a mechanism to evaluate whether the original functionalities of the system have been preserved. In the case of migrating existing code, this aspect is more relevant, as the goal is to improve the performance of the system. In the case of implementing new functions, practitioners must make sure that the newly introduced functionalities have brought a positive impact to the other parts of the system.

We address these aspects through the following question.

\vspace{0.5cm}
\noindent\fbox{%
    \parbox{\linewidth}{%
        Q30: Have the functionalities been preserved despite re-architecting?
    }%
}

\vspace{0.5cm}
\textit{The case study}:
The functionalities been preserved despite refactoring (Q30). Ten thousand input images were inserted into both versions, and the output was identical compared to the original CPU implementation. In this case, only one application was refactored for execution on an accelerator. In larger projects, however, the risk for the functionality not to be preserved in its entirety is higher, requiring more robust testing techniques.

\subsubsection{Determining the outcomes of refactoring}
Re-designing the system brings several additional efforts, but also ideally several additional benefits. On the business level, there must be a calculation of the cost-benefit of the re-architecting efforts with respect to the gains obtained in performance. Such measurements may include energy consumption, the capacity of the resources, and user perception of improvement. The evaluation can also refer to design-time characteristics, such as the measurement of efforts, in order to assess the aspects connected to lifecycle. 

We address these aspects through the following questions.

\vspace{0.5cm}
\noindent\fbox{%
    \parbox{\linewidth}{%
        Q31: Were the expected gains in performance achieved (e.g., execution time, maintainability)? 
        
        Q32: What was the cost or re-architecting?
        
        Q33: What was the trade-off (cost/effort) of re-architecting vs. the gains in performance?
    }%
}

\vspace{0.5cm}
\textit{The case study}:
The performance gain from the refactoring was 31\%, where the original implementation processed a frame in 4.8~ms, and the GPU-based version processed a frame in 3.1~ms. These gains represent a satisfactory performance increase in the context of this project (Q31). The new execution time is comparable with the expectations the stakeholders had prior to starting the refactoring activities. The time invested in refactoring was 7h (Q32). The trade-off between the efforts of refactoring hours and the gains in performance was positive according to the pre-established requirements of the system (Q33).
	
\subsection{Framework Overview}	
The overview of the reasoning framework is shown in Table~\ref{tab:framework}. It includes the stages, the aspects, and the questions that are respective to each aspect. The stages follow the reasoning process considering the different phases of software development for heterogeneous platforms. However, in practice these phases may not necessarily be followed sequentially as shown in the table, but rather suggest a flow that includes revisiting and feedback loops in the process. The proposed stages can be integrated with classical software development processes and therefore assume a complementary role to the reasoning that is typically performed in the development applications, as shown next.

\begin{table*}[ht]
\caption{Overview of the reasoning framework.}
\resizebox{\textwidth}{!}{%
\begin{tabular}{lll}
\hline
\multicolumn{1}{c}{\textbf{Stages}} & \multicolumn{1}{c}{\textbf{Aspects}}                 & \multicolumn{1}{c}{\textbf{Questions}}                                                                                         \\ \hline
\multirow{13}{*}{ASSESSING}         & \multirow{4}{*}{Functionality analysis}              & Q1: Are there bottlenecks in the system?                                                                                       \\ \cline{3-3} 
                                    &                                                      & Q2: Which parts of the system require better performance?                                                                      \\ \cline{3-3} 
                                    &                                                      & Q3: Would the selected function perform more efficiently on a different type of processing unit?                               \\ \cline{3-3} 
                                    &                                                      & Q4: Is the existing software parallelizable?                                                                                   \\ \cline{2-3} 
                                    & \multirow{3}{*}{Hardware decision}                   & Q5: What type of hardware should be used?                                                                                        \\ \cline{3-3} 
                                    &                                                      & Q6: What hardware vendor should be used?                                                                                         \\ \cline{3-3} 
                                    &                                                      & Q7: What vendor-specific toolchain is needed for the change?                                                                   \\ \cline{2-3} 
                                    & \multirow{3}{*}{Impact analysis}                     & Q8: What would be the impact on the existing architecture?                                                                     \\ \cline{3-3} 
                                    &                                                      & Q9: What would be the impact on the system's performance and lifecycle?                                                        \\ \cline{3-3} 
                                    &                                                      & Q10: What are the expected business benefits (time-to-market, market advantage, user perception)?                              \\ \cline{2-3} 
                                    & \multirow{3}{*}{Project feasibility analysis}        & Q11: What would be the cost of the changes?                                                                                    \\ \cline{3-3} 
                                    &                                                      & Q12: What would be the costs to maintain and evolve the functionalities once migrated?                                         \\ \cline{3-3} 
                                    &                                                      & Q13: What would be other consequences of the change regarding non-functional properties?                                       \\ \hline
\multirow{6}{*}{RE-ARCHITECTING}    & \multirow{3}{*}{Software and hardware mapping}       & Q14: Which functionalities will be executed by the accelerator(s)?                                                             \\ \cline{3-3} 
                                    &                                                      & Q15: How do the potential configurations perform?                                                                              \\ \cline{3-3} 
                                    &                                                      & Q16: What will be the mapping between software components and processing units?                                                \\ \cline{2-3} 
                                    & \multirow{2}{*}{Impact on the software architecture} & Q17: What is the impact of the new architecture on the constraints and requirements at run-time?                               \\ \cline{3-3} 
                                    &                                                      & Q18: What is the impact of the new architecture on the constraints and requirements related to lifecycle and business aspects? \\ \cline{2-3} 
                                    & Overall architecture design                          & Q19: What will be the new system architecture in terms of new/changed architectural structure, components, and communication?  \\ \hline
\multirow{5}{*}{DEVELOPING}         & \multirow{3}{*}{Software development process}        & Q20: What are the limitations of the existing tools, frameworks, programming languages, hardware?                              \\ \cline{3-3} 
                                    &                                                      & Q21: What tools, frameworks and programming languages will be used?                                                            \\ \cline{3-3} 
                                    &                                                      & Q22: What changes will affect the development process (development, testing, cooperative development, etc)?                    \\ \cline{2-3} 
                                    & \multirow{2}{*}{Software components refactoring}     & Q23: What type of interface will be defined for the heterogeneous components?                                                  \\ \cline{3-3} 
                                    &                                                      & Q24: How will data be transferred between the platforms?                                                                       \\ \hline
\multirow{5}{*}{DEPLOYING}          & \multirow{4}{*}{Deployment process}                  & Q25: Will deployment be available only in a development phase, or will dynamic deployment be available?                        \\ \cline{3-3} 
                                    &                                                      & Q26: Which constraints exist in the deployment process (timing constraints, resource capacity, and similar)?                   \\ \cline{3-3} 
                                    &                                                      & Q27: Can components be deployed independently, or all components must be (re)deployed as a set?                                \\ \cline{3-3} 
                                    &                                                      & Q28: What are the capabilities of saving data and system status on the target processing units?                                \\ \cline{2-3} 
                                    & Hardware preparation                                 & Q29: What are the preparations needed for execution?                                                                           \\ \hline
\multirow{4}{*}{EVALUATING}         & Functionality testing                                & Q30: Have the functionalities been preserved despite refactoring?                                                              \\ \cline{2-3} 
                                    & \multirow{3}{*}{Refactoring outcomes}                & Q31: Were the expected gains in performance achieved (e.g., execution time, maintainability)?                                  \\ \cline{3-3} 
                                    &                                                      & Q32: What was the cost or refactoring?                                                                                         \\ \cline{3-3} 
                                    &                                                      & Q33: What was the trade-off (cost/effort) of refactoring vs. the gains in performance?                                         \\ \hline
\end{tabular}%
}
\label{tab:framework}
\end{table*}

\section{The Framework in the Software Development Process}
\label{sec:integration}
\begin{figure}[ht]
  \begin{center}
  \includegraphics[width=1\linewidth,trim=0cm 0cm 0cm 0cm]{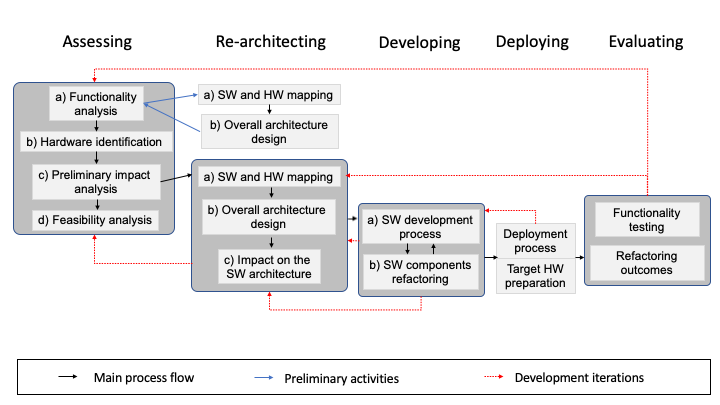}
  \caption{Decision process flow within the framework.}
  \label{fig:frame-flow}
  \end{center}
\end{figure}

Fig~\ref{fig:frame-flow} shows an example of a decision process flow following the decision framework, considering an iterative development process. It shows the main decision process flow that includes the activities that are preliminary to each aspect being addressed. For example, a refactoring feasibility analysis in an early development phase is performed before software and hardware mapping, but a preliminary mapping should still be estimated in order to make the refactoring feasibility analysis enough accurate. The feedback loops between the different aspects are highlighted as development iterations, through which relevant aspects that were previously analyzed are revisited. The figure depicts a typical process in companies using agile or iterative processes. It shows the main flow of the decisions that propagate across the stages (black edges), which in some cases requires preliminary analysis from different stages, e.g., performing a preliminary software and hardware mapping already in the Assessing stage. A typical development process is likely to cycle through several iterations. 

\begin{figure}[ht]
  \begin{center}
  \includegraphics[width=1\linewidth,trim=0cm 0cm 0cm 0cm]{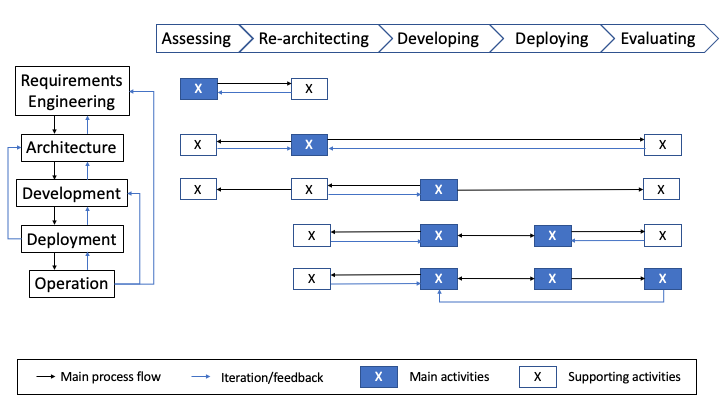}
  \caption{Interaction between the stages in the framework and a typical software engineering process.}
  \label{fig:frame-developmentprocess}
  \end{center}
\end{figure}

Fig~\ref{fig:frame-developmentprocess} shows an example of the decision flow following the decision framework integrated in a development process. The development process is shown as a set of activities (requirements engineering, architecture, development, deployment, and operation) connected with edges that represent an iterative process that is possibly separated in large iterations that include all activities, and smaller iterations in a typical DevOps process~\cite{Lwakatare2016}, which includes continuous development, deployment and evaluation, and possibly system re-architecting. The main reasoning activities integrated in the development process are defined as follows. The assessing stage is tightly coupled with requirements engineering, and it may require re-architecting analysis that might have a large impact on the requirements and constraints. The re-architecting stage includes analysis of hardware reconfiguration and software refactoring. It may require evaluation of the changes not only on the reasoning level, but also with more accurate results, obtained, for example, using simulation and experimentation. The main activities in the developing stage include reasoning about implementation and the actual implementation itself, which is directly connected to the deployment and evaluation stages. In a DevOps process, these activities are strongly interrelated. In the product operation phase, the stages of development, deployment and evaluation are continuously performed, including eventual re-architecting activities.

\section{Evaluation - Feedback from practitioners}
\label{sec:evaluation}

During our research, we have been in close contact with our partner companies in Software Center\footnote{Software Center: https://www.software-center.se}. Particularly, the companies are major players in the automotive domain, with focus on embedded software systems. They are in the process of implementing AI components in their systems and consequently deploying heterogeneous platforms. In total, we presented the framework to six companies that were in different stages in accommodating heterogeneous platforms into their processes. These companies operate in a variety of domains: automotive (passenger and transport), consumer goods, industrial technology, and communications. The participating teams were mostly working on projects in the context of artificial intelligence, thus having computation as one aspect that needs to be addressed. 

We presented the proposed approach and the rationale behind every step in the process. The group then discussed the initiated topic in relation to their day-to-day activities and their own views on how the refactoring process should occur. After several iterations with different partners, we adjusted the framework and sent it back to them for another round of review. The process occurred until all major points of conflicting ideas from the different parties were resolved. Additionally, we conducted interviews with three software architects who are responsible for both research and production projects in two automotive companies and one AI platform company.

With the purpose of evaluating the framework, we sent out a questionnaire\footnote{The questionnaire is available online at https://tinyurl.com/uskxrky} in order to capture the respondents' background information along with their impressions of the framework.
The questionnaire that was sent out contained three questions regarding the professionals' background, experience, and typical software structure in their daily work. Then, we asked several questions that are tightly connected to each aspect/question in the framework. Our goal with this approach was to allow them to elaborate on the feasibility and relevance of the aspects/questions through sharing their own experiences. Finally, the questionnaire also included more open and generic questions about each stage in the framework, with specific focus on what was possibly missing or stated in excess. We received feedback from experts in the following roles: 

\begin{itemize}
\item Head of research \& main architect in an AI platform company, with 20 years of experience, hereby identified as I1;
\item Lead research engineer in an automotive company, with 7 years of experience, hereby identified as I2; and
\item Technology \& strategy leader in an automotive company, with 25 years of experience, hereby identified as I3.
\end{itemize}

The overview of the evaluation results is shown in Table~\ref{tab:triangulation}. It includes a triangulation of sources to demonstrate the framework's usability. The first column presents the stages and the aspects in the framework. The second column includes a summary of the activities performed in the case study with respect to each aspect in the framework. It partially answers the respective questions, mostly focusing on the aspects. This column demonstrates how the aspects in the framework are addressed in the particular case study. The third column, on the other hand, expresses the validity of the aspects and questions in relation to the experts' experiences in developing software in their domains. The individual point-of-view expressed in the text is mapped to the respective interviewee through their IDs in parenthesis. The contents present an overview of reflections of the aspects and questions in the framework with respect to feasibility, relevance, and difficulties surrounding them.

\begin{table}
\caption{Triangulation of sources to evaluate the framework: case study \& questionnaire. \newline}
\tiny
\centering
\hspace*{-1.5cm}\begin{tabular}{p{2.5cm} p{6.5cm} p{6.5cm}}
\begin{tabular}[c]{@{}l@{}}\textbf{Stages / Aspects}\\\end{tabular} & \textbf{Activity performed in the Case Study}                                                                                                                                                                                                                                                                                                             & \textbf{Evaluation of the aspects/questions according to experts}                                                                                                                                                                                                                                                                             \\ 
\hline
ASSESSING                                                           &                                                                                                                                                                                                                                                                                                                                                           &                                                                                                                                                                                                                                                                                                                                                     \\ 
\cline{1-1}
Functionality analysis                                              & The functionality selected in the case study (2D convolution) was underperforming and it was not parallelizable using CPU only, requiring an accelerator to fulfill the performance requirements.                                                                                                                                                         & This aspect is mostly valid for applications in domains such as: 
image processing in the domain of computer vision (I2), and computational physics (I3); in which certain functionalities are very computing intensive and can be designed using several different techniques.       \\
\\
Hardware decision                                                   & Since a computer vision functionality was selected, the operation on each frame can be fully parallelized, therefore a GPU 
is preferred. Such decision also accounted for the developers' focus on deployability and scalability.                                                                & \textit{(Not explicitly asked in the questionnaire - this aspect was added afterwards)}~                                                                                                                                                                                                                                                            \\
\\
Impact analysis                                                     & 
The refactoring included the adoption of the Vulkan / SPIR-V framework on GPU  with possibility to scale up for future changes in the hardware. It is expected that the execution time of the 2D convolution task will decrease.              
& This aspect in particular is difficult to address since it is generally difficult to predict performance on an accelerator (I1). However, it is possible to estimate through code analysis (I2) and worst-case execution time estimation (I3). The impact on the existing architecture largely depends on the project.                              \\
\\
Project feasibility analysis                                        & The project is feasible due to the low investments required for the re-architecting. The selected software standards are open, therefore no special tooling or vendor relations are needed. The developer might need a GPU-programming training. 
& One should not overlook the evaluation of the need for migration, which must be identified (I2). Additional costs should be accounted for tooling, direct migration costs, maintenance, integration and verification (I3). 
\\ 
\hline
RE-ARCHITECTING                                                     &                                                                                                                                                                                                                                                                                                                                                           &                                                                                                                                                                                                                                                                                                                                                     \\ 
\cline{1-1}
Software and hardware mapping                                       & In this case, the CPU fetches, uploads and downloads image frames to and from the GPU, which in turn executes the 2D convolution. The CPU software interacts with the Vulkan API and SPIR-V software that is turned into GPU machine code.                                                                                                                & In some cases, it can be rather straightforward (I1) or require a great level of understanding of the algorithms (I2). Even when the decision is quick, careful system design is needed, particularly when the cost-benefit is more even (I3).                                                                                                      \\
\\
Impact on the software architecture                                 & Most of the impact is related to the memory spaces after refactoring. The edge detector application is executed on the GPU, and the images are stored in the GPU memory space. The decision was made based on the need for faster response and increased maintainability.                                                                                 & In theory, not much should change when using standard solutions. In practice, several aspects must be adjusted (I1), particularly within the application and its interfaces (I2) (I3).                                                                                                                                                              \\
\\
Overall architecture design                                         & Not changed drastically, as the original interfaces are maintained and the existing message passing protocol is preserved. Primary/secondary architectural pattern characteristics are inherently implemented due to the current CPU + GPU design decisions in the frameworks.                                                                            & Ideally, the architectural patterns realizing heterogeneous platforms should focus on dependability and redundancy, encapsulation, domain independence (I3). Alternatively, no particular architectural pattern is used (I2).                                                                                                                       \\ 
\hline
DEVELOPING                                                          &                                                                                                                                                                                                                                                                                                                                                           &                                                                                                                                                                                                                                                                                                                                                     \\ 
\cline{1-1}
Software development process                                        & The project included the Vulkan API, the necessary intermediate language SPIR-V and the programming language C++. The development process has focus on lightweight steps, the generation of binaries, and systematic testing procedures.                                                                                                                  & Both manufacturers' provided and third-party tools are used (I2). Debugging and testing are compromised in the deep learning domain (I1). In terms of processes, the standard, existing procedures are followed (I2), typically agile processes.                                                                                                    \\
\\

Software components refactoring                                     & In this case, only the actual operation was affected and its required memory usage: copying from the shared memory into the GPU memory and initiating the operation. The result is then copied to the shared memory.                                                                                                                                      & In industrial contexts, there must be compliance with the hardware supported API (I2). The communication is enabled through ethernet, CAN, and SPI (I3), therefore the refactoring must comply with these standards' constraints.                                                                                                                   \\ 
\hline
DEPLOYING                                                           &                                                                                                                                                                                                                                                                                                                                                           &                                                                                                                                                                                                                                                                                                                                                     \\ 
\cline{1-1}
Deployment process                                                  & The process is relatively simple, only requiring a GPU machine code to be uploaded using a SPIR-V binary. Limited work memory is needed. Due to the nature of the application, continuous deployment is not yet discussed.                                                                                                                   & The deployment discussion is important because continuous retraining and redeployment is crucial in some cases (I1). Deployment decisions are based on available performance, resources and capabilities (I3).~                                                                                                                                     \\
\\
Hardware preparation                                                & No particular preparations are needed, as the GPU code is generated at runtime.                                                                                                                                                                                                                                                                           & In some cases, deployment decisions require engineering support that must be available (I3). Some domains certainly have more constraints than others.                                                                                                                                                                                              \\ 
\hline
EVALUATING                                                          &                                                                                                                                                                                                                                                                                                                                                           &                                                                                                                                                                                                                                                                                                                                                     \\ 
\cline{1-1}
Functionality testing                                               & A test with 10,000 images was performed with identical results when comparing the CPU implementation with the CPU+GPU implementation. Therefore, the refactoring has successfully preserved the functionalities in the project.~                                                                                                                          & To preserve functionality, it is best when the software is designed with accelerator support initially (I1). Whenever it is not and refactoring is required, there are always discrepancies identified in the procedure, which are solved iteratively and over time (I3).                                                                           \\
\\
Refactoring outcomes                                                & In this case, the adoption of a CPU+GPU implementation allowed for 31\% in performance gains. Such improvement is compatible with the expectations prior to refactoring.                                                                                                                                                                                  & The outcomes are typically measured in development time and hardware costs (I1). They can also be measured in man hours, tools, and equipment (I3).                                                                                                  \\
\hline
\end{tabular}
\label{tab:triangulation}
\end{table}

The evaluation procedure presented in this paper consists of the synthesis of data from multiple sources: industrial contexts in the automotive and AI domain (partner companies), a hybrid academia-industrial environment (case study). Due to this diversity of sources, we have observed aspects that should be accounted for when applying the framework. For instance, not all questions are relevant for every software development context. In particular, we have observed that the impact of re-architecting is widely discussed in literature and agreed upon in industrial discussions, but it might be more relevant on more complex systems. In the case study, it was shown to be an aspect that did not represent great concern in such particular context, using a standardized set of tools and being limited to one functionality. Another example is the software and hardware mapping, which in some cases might be straightforward to selecting the most computationally intensive tasks that are appropriate for execution on a given accelerator. Further, the refactoring of particular software components has shown to be very particular to the project, representing either a simple modification to accommodate a new framework, or, probably in more complex environments, an accountable effort to adapt the existing software.

\subsection{Highlights from the interaction with the experts}

The comments were complementary to the discussions that occurred during the meetings and reflected their opinions based on their experience in the area. The most important comments we received are synthesized and presented as follows.

\textit{Priority to software:} The projects typically put software in the center, i.e., the first step is to implement the functionalities according to predefined requirements, and only then, there is an evaluation of which types of hardware are needed for improved execution. The functionalities that entail applications are the main focus for the development of the systems.

\textit{Continuous refactoring and first version on CPU:} The refactoring is regularly conducted due to constantly changing requirements, despite the high complexity of the projects. Therefore, the process usually starts with a CPU implementation that is migrated to other platforms. It is common that the functionalities to be executed on heterogeneous platforms have already been implemented for execution on CPUs. Then, after analysis and eventual identification of the need for better performance, the refactoring efforts are considered. Therefore, it is relevant that the implementation of new functionalities preferably follows an overall software architecture that allows for such type of migration of code. 

\textit{Migration procedure -  experiments first, deployment later:}  The companies involved in the discussion are in different stages of applying migration of code in their products; not all companies have started with the product development, but are rather in an experimental phase. One of the companies has already adopted a development process that includes code migration, and is using the following process. The CPU-oriented software system (or a subsystem) is used as baseline for performance measurement. Once the algorithm is modified into code that can be executed on an accelerator (e.g., GPU), the performance with emphasis on response time is measured and compared to the performance of the default CPU code. In case the trade-off between performance and other aspects (such as costs) is approved, the developer proceeds to port code to the GPU.

\textit{Importance of simulations:} Simulations have been highlighted in the feedback as a key activity when preparing software for execution on heterogeneous platforms. The common practice is to follow a policy that is heavily based on profiling. Sometimes as early as in the Assessing stage, this activity determines whether the intended migration procedure is feasible, desirable, and whether it is aligned with the company's goals. There are occasions in which the migrated functionalities do not perform as expected, even after a thorough profiling procedure across multiple stages.

\textit{Dependency analysis:} 
Since components are typically developed for execution on CPUs, and CPUs are inherently serial in their execution method, developers must check whether there are any dependencies that prevent algorithms from running in parallel, due to the parallel execution nature of GPUs. When there are no dependencies, the component can more easily be transformed into GPU-runnable code. Otherwise, when there are dependencies, the core functionality within the component must be changed in order to make it parallel. 

\textit{Continuous iterations:} One main aspect is a necessity of iterative approach and feedback loops between  different steps, particularly during the ``software and hardware mapping'' step, in which mechanisms like \textit{profiling} are necessary prior to determining the best configuration according to a given set of requirements. The changes are typically constant and iterative, allowing smaller changes to occur at each iteration.

\section{Threats to validity}
\label{sec:threatstovalidity}
In this section, we present a number of aspects that are relevant to the assessment of this study's validity. These aspects are divided into \textit{internal}, \textit{external}, and \textit{construct} validity.

In terms of internal validity, there is a risk that a portion of relevant literature discussing refactoring of software for heterogeneous platforms was not studied. The body of literature is spread across multiple domains and research communities. We minimized this risk by conducting a systematic mapping study with pre-defined criteria for the selection of studies. Further, the workshops and discussions that were held with practitioners only bring a narrow perspective of the phenomenon in a particular context of their work. We attempted to minimize this threat by including several practitioners with different levels of experience in the events, gathering as many opinions as possible.

The external validity of the study is also addressed by the inclusion of multiple points of view in constructing and adjusting the framework. These considerations were captured from multiple interactions. However, the case study we conducted was restricted to demonstrate the application of the framework using an example in the automotive domain. Some of the steps in the framework were straightforward due to the low complexity of the selected application. We attempted to increase the applicability of the framework by including aspects and questions that are orthogonal to the domain, and at a reasonably high level of granularity. The consequence is that addressing some of the aspects using the framework can be straightforward in some contexts; on the other hand, the framework is applicable to a wider range of situations.

The construct validity is addressed by including in the framework aspects that are not necessarily restricted to the process of refactoring software. Although its application is demonstrated in a process-like format assuming the implementation from CPU-based to heterogeneous platforms, several aspects/questions can be used also in the context of the development of new functionality or even entirely new systems. The stages can also be addressed independently in projects where the goal is not to refactor. One relevant threat is that we might have missed to include all necessary steps to successfully refactor software to heterogeneous platforms across every situation. Adjustments beyond the ones suggested in the framework might be needed.

\section{Related work}
\label{sec:relatedwork}


Challenges in heterogeneous computing have been the subject of many studies in the latest years \cite{Brodtkorb2010}. In previous work, we have analyzed a number of studies in the area of heterogeneous computing published with focus on concerns and approaches that are commonly used in research and practice \cite{Andrade2019,Andrade2019b}. To the best of our knowledge, there are no holistic frameworks that focus on the software engineering lifecycle and support the refactoring of software for heterogeneous platforms. There are, however, some approaches proposing solutions to key challenges that are relevant in this area. The proposals typically focus on the re-architecting efforts, as shown next.

\textit{Software and hardware mapping:}
There are several approaches in literature to support the decision-making process of mapping software and hardware. One of the most relevant is described in \cite{Svogor2013} and \cite{Svogor2019}, in which a genetic algorithm to find a locally optimal solution in respect to a defined cost function, and a model that takes into consideration both the system constraints and user-defined architectural decisions. Further, there are approaches that enable the specification of many requirements or resource constraints, as well as the communication capacity required for interaction between the components, attempting to find a local optimum for a set of components \cite{Svogor2019,Santana2015}. In \cite{Sapienza2013,Sapienza2013b,Sapienza2014}, the authors propose a multi-criteria decision process for hardware/software development based on requirements and constraints related to runtime, lifecycle and business.  

\textit{Architecture design:}
Two key aspects have been addressed when designing the architecture \cite{Andrade2012,Gayen2007}: communication and computation. One way to practically address them is through the \textit{separation of concerns} technique, setting ``communication’’ and ``computation’’ as the main concerns in the center. Components that exchange messages are placed closer in the architecture, as well as components that are executed by the same processor. Another possibility is to establish three main concerns in the architecture: \textit{application model}, \textit{platform model}, and \textit{mapping between application and platform}, as presented in \cite{Andrade2012}. The computation aspect is most commonly addressed through a \textit{master-slave architecture} in order to take advantage of the inherent characteristics of heterogeneous platforms which contain, typically, one main processor (e.g. CPU), and one or more accelerators (e.g., GPU for highly parallelized tasks). Alternatively, a \textit{pipelined architecture} can be implemented, allowing software to be represented as general data flow graphs \cite{Gayen2007}, with particular focus on the performance. The approach bases the allocation strategy on the simulation of executing these graphs.

\textit{Software components refactoring:}
Part of the refactoring process consists of determining \textit{how} the software components will be either designed or refactored. The concept of \textit{flexible software components} can be used in order to create software components that can be executed in any of the available processing units. Support for this type of component design has been proposed earlier in the context of GPUs \cite{Campeanu2017}. Flexible software components allow developers to focus on implementing functions, while mechanisms (namely \textit{adapters}) automatically transfer data between components, taking into consideration the platform specifications. 

\textit{Adoption framework:} A similar concept was found online on Microsoft's website \cite{azure}. The company describes a framework in the form of a step-by-step guide for engineers to adopt their cloud solution: Microsoft Azure. The framework is divided into stages with subtopics and rationale behind each action, though the stages do not contain questions to be answered.

\section{Conclusion \& Future work}
\label{sec:conclusion}

Heterogeneous computing has emerged as the most important way of achieving high levels of performance in domains such as embedded systems. It allows software to be executed by processing units with fundamentally different characteristics, often allowing for better performance results and energy efficiency when compared to standard procedural CPU-focused implementations. However, the process of refactoring software for execution on different processing units might be challenging due to a number of aspects, such as the often-significant increase in complexity of the software. Further, there are constraints and requirements that must be fulfilled in order to make the results worth it despite the extra engineering effort. To the best of our knowledge, a holistic approach that guides engineers in the process of performing refactoring of software systems has not yet been proposed. 

In this paper, we proposed a decision framework that aids engineers in the process of refactoring software systems when migrating from CPU-based projects to execution on heterogeneous platforms. The framework is divided into 5 stages that should be followed, not necessarily sequentially, in order to reflect upon different aspects that are relevant for the task. The stages are, namely: \textit{``Assessing''}; \textit{``Re-architecting''}; \textit{``Developing''}; \textit{``Deploying''}; and \textit{``Evaluating''}. The proposed aspects are connected to a typical software engineering process that allows for feedback loops and re-visitation of other stages.

The proposed framework is the result of previous studies in which we thoroughly reviewed the literature on the topic and empirically identified the challenges that practitioners currently face in industrial contexts. Both perspectives have culminated in the creation of this approach, which was presented to---and validated with---multiple industrial partners in the form of meetings and workshops. The relevance of the approach was attested through a set of interviews and a questionnaire that was sent out in order to obtain feedback from these practitioners. The applicability of the approach was demonstrated through a case study in the automotive domain, in which we implemented an edge detector as a 2D convolution operation.

The main goal of the proposed decision framework is to guide practitioners (engineers, architects, analysts) in their decision process to migrate software to heterogeneous platforms. We have particularly focused on the cyber-physical systems domain, which is characterised by distributed computing in real time, with increased demands on computational resources. In other domains, the decision process could be different, depending to a large extent on the available technology. Typical examples are on-line services running AI algorithms. 

Further, we have observed that good practice in the adoption of heterogeneous computing typically translated into incremental engineering, starting with experimentation, simulation, and continuous integration and deployment of parts of the system. 

As future work, we intend to further refine the proposed approach through case studies in other domains in order to capture aspects that might have been missed in the current version. This includes reasoning upon differences across different domains, the deployment process itself, and integration with specific, industrial development processes. In particular, the introduction of Machine Learning and AI components poses additional requirements in distinguishing a development environment with training procedures from the runtime environment.

\section*{Funding}
This research was supported by the research projects ``HELPING -- Heterogeneous Platform Deployment Modelling of Embedded Systems'' funded by the Swedish Research Council, and ``HoliDev -- Holistic DevOps Framework'' funded by Vinnova.



\bibliographystyle{elsarticle-num} 
\bibliography{library}

\begin{thebibliography}{10}
\expandafter\ifx\csname url\endcsname\relax
  \def\url#1{\texttt{#1}}\fi
\expandafter\ifx\csname urlprefix\endcsname\relax\def\urlprefix{URL }\fi
\expandafter\ifx\csname href\endcsname\relax
  \def\href#1#2{#2} \def\path#1{#1}\fi

\bibitem{Chen2017}
Y.~{Chen}, T.~{Krishna}, J.~S. {Emer}, V.~{Sze}, Eyeriss: An energy-efficient
  reconfigurable accelerator for deep convolutional neural networks, IEEE
  Journal of Solid-State Circuits 52~(1) (2017) 127--138.

\bibitem{Garland2008}
M.~{Garland}, S.~{Le Grand}, J.~{Nickolls}, J.~{Anderson}, J.~{Hardwick},
  S.~{Morton}, E.~{Phillips}, Y.~{Zhang}, V.~{Volkov}, Parallel computing
  experiences with cuda, IEEE Micro 28~(4) (2008) 13--27.
\newblock \href {https://doi.org/10.1109/MM.2008.57}
  {\path{doi:10.1109/MM.2008.57}}.

\bibitem{CUDA}
\href{https://developer.nvidia.com/}{{CUDA} architecture overview} (2020).
\newline\urlprefix\url{https://developer.nvidia.com/}

\bibitem{vulkan}
\href{https://www.khronos.org/vulkan/}{Vulkan -- the new generation graphics
  and compute api} (2020).
\newline\urlprefix\url{https://www.khronos.org/vulkan/}

\bibitem{AUTOSAR}
{AUTOSAR}, http://autosar.org, \textit{Accessed 2020-09-16} (2002).

\bibitem{Dresch2014}
A.~Dresch, D.~P. Lacerda, J.~A.~V. Antunes, Design Science Research: A Method
  for Science and Technology Advancement, Springer Publishing Company,
  Incorporated, 2014.

\bibitem{Vaishnavi2007}
V.~K. Vaishnavi, W.~Kuechler, Jr., Design Science Research Methods and
  Patterns: Innovating Information and Communication Technology, 1st Edition,
  Auerbach Publications, Boston, MA, USA, 2007.

\bibitem{Andrade2019}
H.~{Andrade}, J.~{Schroeder}, I.~{Crnkovic}, Software deployment on
  heterogeneous platforms: A systematic mapping study, IEEE Transactions on
  Software Engineering (2019) 1--1\href
  {https://doi.org/10.1109/TSE.2019.2932665}
  {\path{doi:10.1109/TSE.2019.2932665}}.

\bibitem{Andrade2018}
H.~{Andrade}, I.~{Crnkovic}, A review on software architectures for
  heterogeneous platforms, in: 2018 25th Asia-Pacific Software Engineering
  Conference (APSEC), 2018, pp. 209--218.
\newblock \href {https://doi.org/10.1109/APSEC.2018.00035}
  {\path{doi:10.1109/APSEC.2018.00035}}.

\bibitem{Andrade2019b}
H.~{Andrade}, L.~E. {Lwakatare}, I.~{Crnkovic}, J.~{Bosch}, Software challenges
  in heterogeneous computing: A multiple case study in industry, in: 2019 45th
  Euromicro Conference on Software Engineering and Advanced Applications
  (SEAA), 2019, pp. 148--155.
\newblock \href {https://doi.org/10.1109/SEAA.2019.00031}
  {\path{doi:10.1109/SEAA.2019.00031}}.

\bibitem{opendlv}
\href{https://github.com/chalmers-revere/opendlv}{Open{DLV}}, \textit{Accessed
  2020-09-10} (2020).
\newline\urlprefix\url{https://github.com/chalmers-revere/opendlv}

\bibitem{Revere}
\href{https://www.chalmers.se/en/researchinfrastructure/revere}{{{REVERE} -
  Research Vehicle Resource}}, \textit{Accessed 2020-09-10} (2020).
\newline\urlprefix\url{https://www.chalmers.se/en/researchinfrastructure/revere}

\bibitem{Docker}
\href{https://www.docker.com}{{Docker}}, \textit{Accessed 2020-09-10} (2020).
\newline\urlprefix\url{https://www.docker.com}

\bibitem{libcluon}
\href{https://github.com/chrberger/libcluon}{libcluon}, \textit{Accessed
  2020-09-10} (2020).
\newline\urlprefix\url{https://github.com/chrberger/libcluon}

\bibitem{CaseStudyRepo}
{Case study repository: Convolution using {CPU} and {GPU} in Open{DLV}},
  https://gitlab.com/olbender/hpm-paper-casestudy, \textit{Accessed 2020-10-14}
  (2020).

\bibitem{Andrade2020}
H.~{Andrade}, C.~{Berger}, I.~{Crnkovic}, J.~{Bosch}, Principles for
  re-architecting software for heterogeneous platforms, in: 2020 27th
  Asia-Pacific Software Engineering Conference (APSEC), 2020, pp. 1--1.

\bibitem{Lwakatare2016}
L.~E. Lwakatare, P.~Kuvaja, M.~Oivo, An exploratory study of devops extending
  the dimensions of devops with practices, in: ICSEA 2016 : The Eleventh
  International Conference on Software Engineering Advances, 2016, pp. 91--99.

\bibitem{Brodtkorb2010}
A.~R. Brodtkorb, C.~Dyken, T.~R. Hagen, J.~M. Hjelmervik, O.~O. Storaasli,
  \href{http://dx.doi.org/10.1155/2010/540159}{State-of-the-art in
  heterogeneous computing}, Sci. Program. 18~(1) (2010) 1--33.
\newblock \href {https://doi.org/10.1155/2010/540159}
  {\path{doi:10.1155/2010/540159}}.
\newline\urlprefix\url{http://dx.doi.org/10.1155/2010/540159}

\bibitem{Svogor2013}
I.~Švogor, I.~Crnković, N.~Vrček, Multi-criteria software component
  allocation on a heterogeneous platform, in: Proceedings of the ITI 2013 35th
  International Conference on Information Technology Interfaces, 2013, pp.
  341--346.
\newblock \href {https://doi.org/10.2498/iti.2013.0558}
  {\path{doi:10.2498/iti.2013.0558}}.

\bibitem{Svogor2019}
I.~Švogor, I.~Crnković, N.~Vrček, An extensible framework for software
  configuration optimization on heterogeneous computing systems: Time and
  energy case study, Information and Software Technology 105 (2019) 30 -- 42.
\newblock \href {https://doi.org/https://doi.org/10.1016/j.infsof.2018.08.003}
  {\path{doi:https://doi.org/10.1016/j.infsof.2018.08.003}}.

\bibitem{Santana2015}
L.~Sant'Ana, D.~Cordeiro, R.~Camargo, Plb-hec: A profile-based load-balancing
  algorithm for heterogeneous cpu-gpu clusters, in: 2015 IEEE International
  Conference on Cluster Computing, 2015, pp. 96--105.
\newblock \href {https://doi.org/10.1109/CLUSTER.2015.24}
  {\path{doi:10.1109/CLUSTER.2015.24}}.

\bibitem{Sapienza2013}
G.~{Sapienza}, T.~{Secelanu}, I.~{Crnkovic}, Modelling for hardware and
  software partitioning based on multiple properties, in: 2013 39th Euromicro
  Conference on Software Engineering and Advanced Applications, 2013, pp.
  189--194.

\bibitem{Sapienza2013b}
G.~{Sapienza}, T.~{Seceleanu}, I.~{Crnknovic}, Partitioning decision process
  for embedded hardware and software deployment, in: 2013 IEEE 37th Annual
  Computer Software and Applications Conference Workshops, 2013, pp. 674--680.

\bibitem{Sapienza2014}
G.~{Sapienza}, I.~{Crnkovic}, P.~{Potena}, Architectural decisions for hw/sw
  partitioning based on multiple extra-functional properties, in: 2014
  IEEE/IFIP Conference on Software Architecture, 2014, pp. 175--184.
\newblock \href {https://doi.org/10.1109/WICSA.2014.19}
  {\path{doi:10.1109/WICSA.2014.19}}.

\bibitem{Andrade2012}
H.~A. Andrade, A.~Ghosal, K.~Ravindran, B.~L. Evans, A methodology for the
  design and deployment of reliable systems on heterogeneous platforms, in:
  2012 International Conference on Reconfigurable Computing and FPGAs, 2012,
  pp. 1--7.
\newblock \href {https://doi.org/10.1109/ReConFig.2012.6416722}
  {\path{doi:10.1109/ReConFig.2012.6416722}}.

\bibitem{Gayen2007}
S.~Gayen, E.~J. Tyson, M.~A. Franklin, R.~D. Chamberlain,
  \href{http://dx.doi.org/10.1109/PADS.2007.5}{A federated simulation
  environment for hybrid systems}, in: Proceedings of the 21st International
  Workshop on Principles of Advanced and Distributed Simulation, PADS '07, IEEE
  Computer Society, Washington, DC, USA, 2007, pp. 198--210.
\newblock \href {https://doi.org/10.1109/PADS.2007.5}
  {\path{doi:10.1109/PADS.2007.5}}.
\newline\urlprefix\url{http://dx.doi.org/10.1109/PADS.2007.5}

\bibitem{Campeanu2017}
G.~{Campeanu}, J.~{Carlson}, S.~{Sentilles}, Flexible components for
  development of embedded systems with gpus, in: 2017 24th Asia-Pacific
  Software Engineering Conference (APSEC), 2017, pp. 219--228.
\newblock \href {https://doi.org/10.1109/APSEC.2017.28}
  {\path{doi:10.1109/APSEC.2017.28}}.

\bibitem{azure}
\href{https://docs.microsoft.com/en-us/azure/cloud-adoption-framework/}{{Microsoft
  Cloud Adoption Framework for Azure}}, \textit{Accessed 2020-10-18} (2020).
\newline\urlprefix\url{https://docs.microsoft.com/en-us/azure/cloud-adoption-framework/}

\end{thebibliography}





\end{document}